\documentclass[runningheads]{llncs}
\usepackage{xspace}
\usepackage{xcolor}
\usepackage{url}

\usepackage{subfig}
\usepackage{makecell}
\usepackage[T1]{fontenc}

\usepackage{graphicx}
\newcommand{\mr}[1]{#1}
\newenvironment{mrlong}{}{}

\begin{document}
\title{Quantifying User Password Exposure to Third-Party CDNs}
\author{Rui Xin, Shihan Lin, Xiaowei Yang}
\authorrunning{Rui Xin, Shihan Lin, Xiaowei Yang}
\institute{Duke University\\
    \email{rui.xin926@duke.edu, shihan.lin@duke.edu, xwy@cs.duke.edu}
}
\maketitle              % typeset the header of the contribution
\begin{abstract}
    Web services commonly employ Content Distribution Networks (CDNs) for performance and security. As web traffic is becoming 100\% HTTPS, more and more websites allow CDNs to terminate their HTTPS connections. This practice may expose a website's user sensitive information such as a user's login password to a third-party CDN. In this paper, we measure and quantify the extent of user password exposure to third-party CDNs. We find that among Alexa top 50K websites, at least 12,451 of them use CDNs and contain user login entrances. Among those websites, 33\% of them expose users' passwords to the CDNs, and a popular CDN may observe passwords from more than 40\% of its customers. This result suggests that if a CDN infrastructure has a vulnerability or an insider attack, many users' accounts will be at risk. \mr{If we assume the attacker is a passive eavesdropper, a website can avoid this vulnerability by encrypting users' passwords in HTTPS connections.} Our measurement shows that less than 17\% of the websites adopt this countermeasure.
    % \keywords{HTTPS \and CDN \and Password \and Security \and Measurement}
\end{abstract}

\newcommand{\Name}{InviCloak\xspace}
\newcommand{\name}{InviCloak\xspace}
\newcommand{\xy}[1]{\textcolor{blue}{[XY: #1]}}
\newcommand{\lsh}[1]{\textcolor{red}{[#1]}}
\newcommand{\fp}{\vspace*{0.05in}\noindent}
\newcommand{\eg}{{\it e.g.}\xspace}
\newcommand{\ie}{{\it i.e.}\xspace}
\newcommand{\et}{{\it et~al.}\xspace}
\newcommand{\dos}{DDoS\xspace}

\section{Introduction}
Content Distribution Networks (CDNs)~\cite{Akamai-SGICOMMCCR15,Akamai-SIGOPS10}
play an important role in improving the performance and security of web
services. A CDN caches web pages at servers near end users to reduce retrieval latency. It also blocks malicious requests to defend a web server
against various attacks~\cite{Protecting-Computer15}. Currently, many websites employ CDNs provided by third-party companies such as Akamai~\cite{Akamai}, Cloudflare~\cite{Cloudflare}, and Fastly~\cite{Fastly}. 

However, third-party CDNs introduce a considerable security and
privacy risk when they serve websites that enable HTTPS~\cite{KeySharing-CCS16,Interception-NDSS17}. HTTPS uses a certificate to certify the domain name of a website. Thus, to make the web pages appear as if they come from the
original site, a website has to share its TLS private key \cite{KeySharing-CCS16} or TLS session keys\cite{CloudflareKeylessSSL} with the CDN. In both cases, a third-party CDN can observe the content of all connections between a website and its users.

 In this work, we aim to raise awareness of this security and privacy risk and quantify its severeness from a user's perspective. We choose to measure the extent to which users' website login passwords are exposed to CDNs due to the HTTPS key sharing practice. 
 Although prior research has shown that private key sharing is prevalent on the Internet ~\cite{KeySharing-CCS16} and HTTPS termination weakens connection
security of a great portion of the Internet~\cite{Interception-NDSS17}, 
it is not clear whether websites have taken preliminary countermeasures such as 
client-side encryption (see \S~\ref{Background}) to protect users' passwords in the case of a passive attacker.
% Passwords guard users' private information stored at a website. If an attacker obtains a user's password at one website, not only can he login as the user at that website, but also he may launch credential stuffing attacks~\cite{xxx} to login as the
% user at other websites. This is because users tend to re-use their
% passwords across websites~\cite{xxx}. 

% Prior research presents that private key sharing is prevalent on the Internet
% ~\cite{KeySharing-CCS16} and HTTPS termination weakens connection security of a great portion of the Internet~\cite{Interception-NDSS17}. However, the security implication for
% users' privacy caused by prevalent HTTPS termination remains unclear. Moreover,
% given the private key sharing practice, a website may take countermeasures such
% as CDN bypassing and client-side encryption (see \S~\ref{Background}) to protect
% users' private data. The deployment of such countermeasures on websites is also
% unclear. Therefore, the research on private data leakage to CDNs from a user's
% perspective remains vacant.

We conduct a measurement on Alexa top 50K sites~\cite{AlexaTopSites} 
to quantify password exposure to CDNs during the
user login procedures. We also measure the deployment of client-side password encryption
on websites to understand websites' treatment of users' passwords. Such a
large-scale measurement is technically non-trivial, because we need
to automate the login procedures on websites with diverse structures to inspect
login requests. Thus, we design and implement a framework for automatic login. The
framework can detect login elements on a website and collect login
requests when it submits credentials to websites. 

Our main contributions and findings can be concluded as the following:
\begin{itemize}
	\item We propose an open-source framework for automatic login~\footnote{The code is available at \url{https://github.com/SHiftLin/PAM2023-CDNPassword}}, which can be
	applied to other research such as the measurement of authentication methods.
	\item Our measurement presents that 33.0\% of websites that employ CDNs and
	contain login entrances expose users' passwords in plaintext to their CDNs. 
	\item We find that two popular CDN providers, Cloudflare and Akamai, can
	observe users' passwords from 44\% and 25\% of their customers,
	respectively.
	\item \mr{We find prevalent password exposure in most website categories,
	including websites whose user accounts should be carefully protected,
	such as websites related to finance and health. Retail websites substantially benefit from CDNs,
        but most of them (58\%) expose passwords to CDNs.}
	\item Our result shows that less than 17\% of the websites encrypt users' passwords
	when transferring login requests to CDNs, and the top 1,500 websites are more likely to adopt client-side password encryption.
\end{itemize}

Overall, our measurement points out potential security issues caused by password
exposure to CDNs. Even though websites trust CDNs, users may concern
about their privacy when CDNs can monitor their private data including passwords. Moreover, CDNs have never been secure enough. Prior work has shown that an attacker can trick some CDNs to cache and reveal other users'
private data~\cite{WCD,WCD-Security20,WCD-Security22}. Thus, private data leakage to
CDNs may turn into a disaster when attackers or malicious insiders exploit vulnerabilities of CDNs.

\section{Background}~\label{Background} In this section, we briefly introduce
CDNs and HTTPS, and we analyze the security issues when a website with HTTPS
employs a CDN. We also discuss two countermeasures adopted by websites in practice to
address such issues.

\subsection{HTTPS on CDNs}
A CDN reduces web retrieval time by directing a client's request to an \emph{edge
server} which is hosted by the CDN and geographically close to the user. The edge server responds to
the client with cached content. If the requested content is not cached, the edge
server may fetch the content from the \emph{origin server} which is hosted by
the website (the CDN's \emph{customer}) and is the initial source of all
content. CDNs do not cache private data, as they are usually dynamic. 

Modern CDNs are used not only to speed up page loading but also to provide an
effective shield against attacks such as DDoS and code
injections~\cite{Protecting-Computer15}. A CDN enlarges the serving capability of
its customers to prevent volumetric DDoS attacks. It also applies techniques such
as IP blocking and rate limiting to block attacks when DDoS happens. For example,
Akamai protected its customers from 38,905 separate DDoS attacks from 2014 to
2019~\cite{Akamai5YearsDDoS}. CDNs also inspect the content of requests and
use Web Application Firewall (WAF) to filter out malicious requests such as XSS
injection~\cite{XSS-ESORICS11} and SQL injection~\cite{SQLInjection-ISSSE06}. 

Unfortunately, CDNs have become a source of vulnerabilities in the HTTPS
ecosystem in recent years~\cite{KeySharing-CCS16,Interception-NDSS17}.
% The security of HTTPS relies on certificates and private keys generated 
% by totally trusted certification authorities (CAs). 
% However, since HTTPS requires server authentication by
% private keys in HTTPS handshakes, 
If a website employs a CDN to represent it to
respond to clients' HTTPS requests, it has to share its private key with the CDN.
With the private key, the CDN can build HTTPS connections with clients, and
clients cannot differentiate between the CDN and the origin server. When a
client requests for private data, the CDN will forward the request by terminating
the HTTPS connection and building another HTTPS connection with the origin
server. Therefore, the CDN becomes a man in the middle when a user's private
data are transmitted between the client and the origin server~\cite{KeySharing-CCS16}. 

\subsection{Countermeasures in Practice}~\label{ExistingSolutions} Two instant but
imperfect countermeasures have been deployed by some websites. First, a website can
bypass the CDN and send the private requests to the origin server directly. In
this countermeasure, a website should use a separate domain or subdomain for the
private data, because the CDN possesses the private key of the original domain.
We refer to this method as ``\emph{CDN bypassing}'' in this paper.
This method will not affect CDNs' benefit of page loading acceleration, since
the private data are not cached by CDNs. However, it eliminates the benefit of having
the origin server shielded against DDoS attacks, because the IP address of the
origin server is exposed to the public. When attackers can connect to the origin
server directly, it is much easier to launch DDoS attacks since the origin
server usually cannot construct a DDoS defense as effectively as
CDNs~\cite{CDNAbuse-SRDS18,BypassCBSP-CCS15}. Besides DDoS, the CDN cannot
inspect the private content to filter out malicious requests, and thus the
origin server may suffer from attacks such as code injections.

Another countermeasure is to encrypt private data inside HTTPS connections. The website
generates another key pair and delivers the new public key to the client. The
client uses the public key to encrypt the private data to be sent out.
Therefore, when a CDN forwards the request, the private data are invisible to
the CDN. We refer to this method as ``\emph{client-side encryption}'' in our paper. 
We observe some websites use this method to protect users' passwords
only, as encrypting all private data may introduce too much overhead. \mr{However, the client-side encryption only defends against a passive attacker as described in \S~\ref{ThreatModel}. }
Besides, secure public key delivery is non-trivial when HTTPS connections are already
intercepted by a CDN~\cite{InviCloak-CCS22}. 
Delivering another certificate differing from the HTTPS
certificate is useless, because a website has to use JavaScript to conduct
encryption in current browsers, and the JavaScript code cannot obtain the root
certificates of a client to verify a certificate. \mr{Without a certificate, if the
public key is delivered by a CDN, a CDN with an active attacker (defined in \S~\ref{ThreatModel}) inside can launch the
man-in-the-middle attacks by replacing the public key.} If the public key is
delivered by the origin server, the origin server is exposed to the public and
under the threat of DDoS. \mr{In practice, websites use an asynchronous JavaScript call~\cite{AJAX} to request for a public key from the origin server and encrypt passwords by JavaScript code.}

Despite the defects of these two methods, they preserve users' privacy to
some extent. Moreover, if the origin server builds its own DDoS defense or a CDN
is assumed to be a passive attacker, these two countermeasures can
provide sufficient protection. However, it is unclear about the deployment of
these two countermeasures on websites. Thus, we investigate the password exposure to provide a profile of their deployment. 
%Our measurement will show that few websites adopt the client-side encryption for passwords.

\section{Threat Model}~\label{ThreatModel} 
We use the threat model proposed by the prior work~\cite{InviCloak-CCS22}.
We consider the private data in a website as the data can only be accessed by a authenticated user.
The users can be authenticated by the traditional password, one-time password (OTP), OAuth~\cite{RFC6749-OAuth}, certificates, etc.
The credentials for authentication are considered as private data as well.
We focus on the measurement of the traditional password in this paper.

We considered two types of attackers defined in the prior work~\cite{InviCloak-CCS22}.
\begin{itemize}
    \item \textbf{Passive attacker:} A CDN behaves honestly
    to serve the requests, but an attacker inside the CDN may eavesdrop on the transmitted messages. 
    For example, a malicious administer of a CDN cannot change the CDN's behavior
    but may peek at the transmitted traffic and record users' passwords.
    \mr{Client-side encryption can protect users' password under a passive attacker.}
    \item \textbf{Active attacker:} An attacker insider CDN may launch arbitrary
    attacks including eavesdropping and tampering. \mr{Thus, it is more capable than a passive attacker.} For example, a CDN may modify or corrupt the cached HTML or JavaScript
    to disable the client-side encryption so that it can observe users' passwords in the login requests.
    This may happen when attackers exploit a vulnerability of a CDN. 
    \mr{As previously mentioned, CDN-bypassing can defend against an active attacker inside a CDN, but it introduces the vulnerability of DDoS to the origin server.}
\end{itemize}

\section{Method}~\label{Method}
To detect the password exposure, we should inspect a website's login
request and the destination.
Thus, we need a framework for automatic login in a large-scale measurement.
Currently, a website may adopt multiple authentication methods, such as text passwords,
OAuth~\cite{RFC6749-OAuth}, one-time password~(OTP). In our measurement, we only
consider the method of text passwords.

Based on the existing frameworks~\cite{Phishing-AsiaCCS2019,AutoLogin-NDSSWorkshop20,LeakyForm-Security22}, 
we designed and implemented an automatic login framework that copes with more web pages with diverse structures. In our work, we do not need to successfully log into a website, so the framework merely triggers a failed login and collects the login request. \mr{We elaborate on the design and implementation of such a framework in the appendix.}

% Fathom~\cite{LeakyForm-Security22} is 
% a framework based on supervised learning, while our framework is rule-based. 
% We will show that our framework can detect more login forms and perform more successful logins 
% in websites than Fathom in \S~\ref{RankingDistribution}.

%We also differentiate login forms and sign up forms through classifying elements,
%which enables us to handle sign up forms with only two input elements.

Besides the automatic login framework, we use the method in the prior work~\cite{KeySharing-CCS16,CDN-IMC08,CDN-IMC01,CDNMeasure-BSThesis2017,InviCloak-CCS22} to discover the CDN usage of a website.
This method also helps to inspect the destination of the collected login requests to determine whether the requests are sent to a CDN server.

Some cloud providers will provide both hosting service and CDN service,
such as AWS and Azure. In our method, when a request is sent to such a cloud
provider, we cannot determine whether the website is using the CDN service or
the hosting service. If the password is sent to a hosting service, it should
not be considered as an exposure to a CDN. Since our goal is to provide an
underestimation of password exposure, our CDN list does not include a CDN service provider
that also provides hosting service. \mr{As a result, our CDN list contains 9 popular CDNs, namely
    Cloudflare, Akamai, Fastly, Highwinds, Edgecast, Incapsula, Quantil, CDNetworks, and Limelight.}

\mr{We collected 50k websites from the Alexa ranking list~\cite{AlexaTopSites} and ran our experiments of automatic login and CDN discovery in Oct. 2020. In our future work, we will set up our experiments as a monitoring platform to observe the evolution of password exposure behavior.}

\emph{Ethical concerns:} We respect user privacy, and our work does not raise ethical concerns.
The method of CDN discovery only used public data from the Internet, such as Registration Data Access Protocol (RDAP)~\cite{RFC7482-RDAP}.
As for the automatic login framework, since we do not require a successful login, we use a
randomly generated fake account that is nearly impossible to coincide with existing
ones. We skip the websites that require a test of the account existence before
submitting the login credentials. We only conduct the login trial once for each website, so
we do not overload the websites in our test.

\begin{figure*}[tbp]
    \centering
    \subfloat[\textbf{CDF}]{
        \begin{minipage}[t]{0.48\textwidth}
            \centering
            \includegraphics[width=\textwidth]{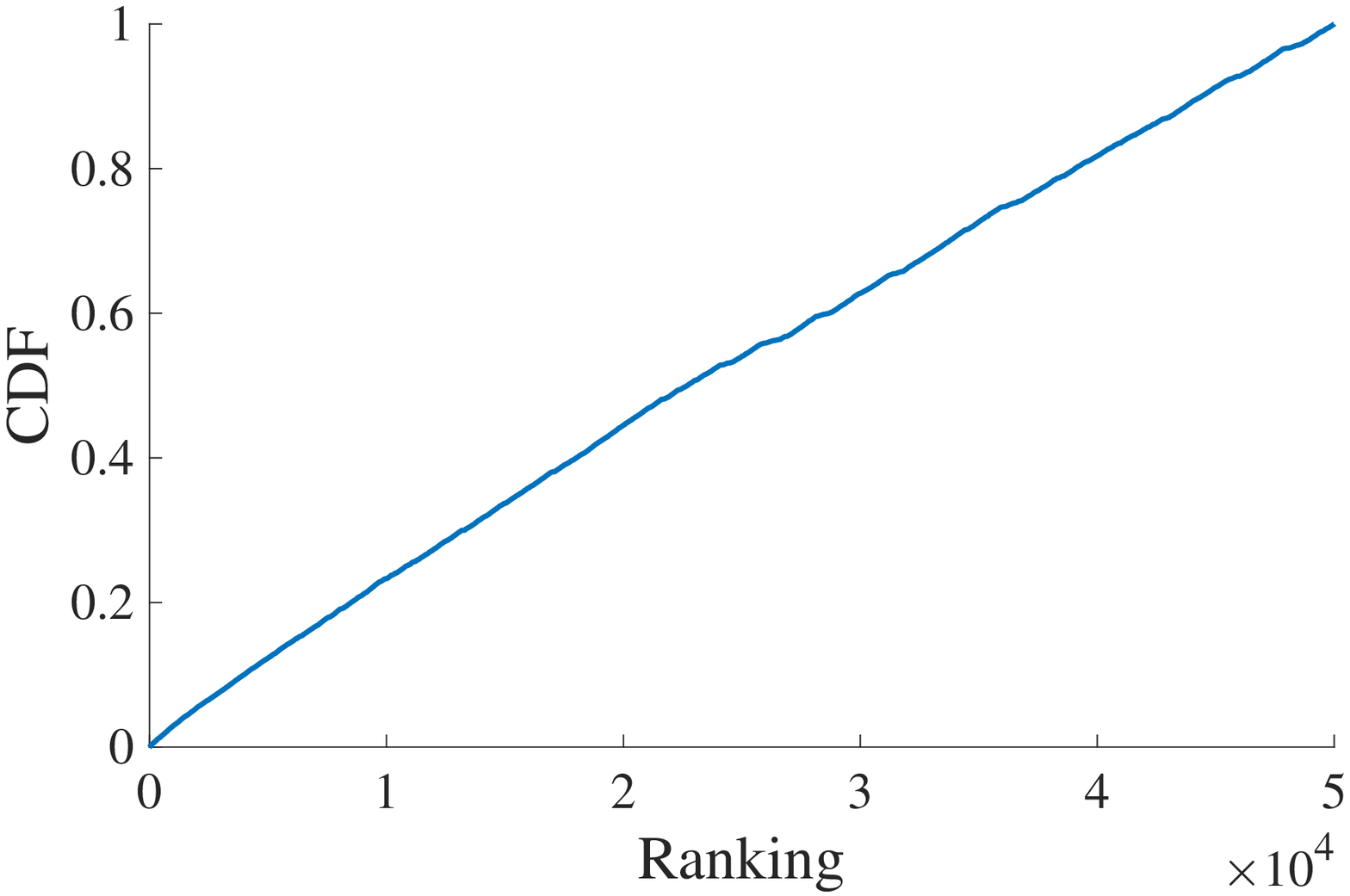}
        \end{minipage}\label{fig:LoginCDF}
    }
    \subfloat[\textbf{Percentage}]{
        \begin{minipage}[t]{0.48\textwidth}
            \centering
            \includegraphics[width=\textwidth]{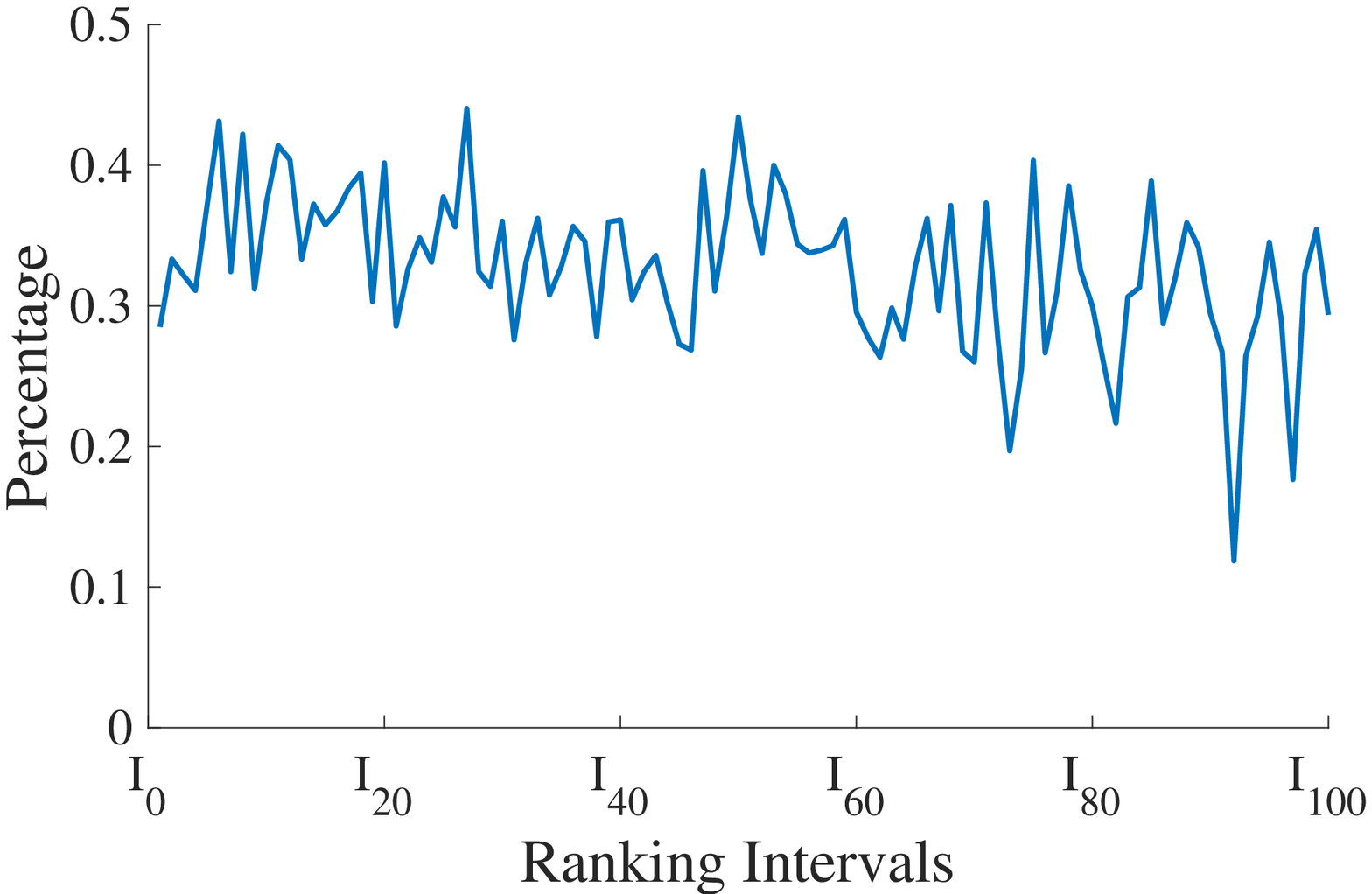}
        \end{minipage}\label{fig:SharePercent}
    }
    \caption{(a) Distribution of login-detected websites. (b) Percentages of password-exposed websites among CDN-enabled websites across different ranking intervals. We divide 50K websites into 100 ranking intervals. Each interval contains 500 websites. The x-axis ticks at every 20 intervals.}
\end{figure*}

\section{Password Exposure}~\label{PasswordExposure}
We only consider HTTPS-enabled websites because a website without HTTPS
apparently contains major vulnerabilities. In Alexa top 50K
sites~\cite{AlexaTopSites}, 42,502 of them enable HTTPS.  We run the framework
to automatically log into these websites. If the framework submits the
fake credentials to a website, we consider it performs a login. The framework
performs 17,111 logins in total.  In this paper, we focus on these 17,111
websites and call them ``login-detected websites''.

We detect CDNs employed by these websites according to
\S~\ref{Method}. Our result shows that 12,451 websites employ CDN service,
and we call them ``CDN-enabled websites'' in this paper. By inspecting their
login procedures, we find that 4,114 websites send the login requests with users'
passwords in plaintext or Base64 encoding to CDNs. We denote these websites as ``password-exposed
websites''. We discovered that 33\% of CDN-enabled websites expose users'
passwords to CDNs, demonstrating a potential privacy issue. In this section, we
present the results in detail.

\subsection{Distribution over Rankings}~\label{RankingDistribution}
Since our framework may fail to detect the login forms of some websites, the dataset
of login-detected websites is a sample set of all websites
that enable logins. We first investigate the distribution of these samples over
rankings.

Figure~\ref{fig:LoginCDF} shows the distribution of login-detected websites. A
linear relationship between the CDF and ranking shows a uniform distribution of the
websites. Therefore, the logins detected by our framework are unbiased in the
rankings.

To investigate the relationship between websites' rankings and their preference for
password exposure, we divide the rankings into 100 intervals. For an interval
$I_j$, it contains 500 websites ranking in the range of $[1+500*(j-1),500*j]$.
For each interval, we count the password-exposed websites and the CDN-enabled
websites, and we compute the percentage of password-exposed websites in
CDN-enabled websites.

Figure~\ref{fig:SharePercent} presents the percentage variation across the
intervals. Given the result of unbiased detection in Figure~\ref{fig:LoginCDF},
we can examine the distribution of password exposure on website rankings
through Figure~\ref{fig:SharePercent}. Even though some fluctuations exist, the
percentages are overall above 20\%, meaning that the password exposure is
common across all rankings. 
% We can also find that the percentages of
% password-exposed websites with a higher ranking are slightly larger than
% those with a lower ranking. The average percentage from the $I_1$ to $I_{50}$
% and from the $I_{51}$ to $I_{100}$ are 34.6\% and 30.8\%, respectively. The
% reason for the difference may be that websites with higher rankings should handle
% more traffic, and they have a stronger preference for adopting CDNs to filter out
% the malicious requests. Thus, those websites tend to expose the private requests
% destined at the origin server to CDNs for inspection. 
Besides, we can find that
the most popular websites in the first two intervals have relatively low password
exposure percentage. It is because that the top websites are more likely to
deploy defense mechanisms, which can be justified by our analysis in
\S~\ref{Countermeasures}.

\begin{table}[t]
    \caption{\mr{Distribution across CDN providers (a) and website categories(b). The ``Percent'' column denotes the percentage of password-exposed websites in CDN-enabled websites. We mark notable data with red color.}}
    \subfloat[\label{tab:CDNDist}\textbf{CDN providers}]{
        \centering
        \begin{tabular}{>{\bfseries\scriptsize}l|c|c|c}
            \hline
            \scriptsize{\textbf{\makecell{CDN            \\ provider}}} & \scriptsize{\textbf{\makecell{CDN-              \\enabled}}} & \scriptsize{\textbf{\makecell{Password-\\exposed}}} & \scriptsize{\textbf{Percent}} \\
            \hline
            Cloudflare & 6356 & 2803 & \color{red}{44}\% \\
            Akamai     & 3280 & 818  & 25\%              \\
            Fastly     & 1631 & 291  & 18\%              \\
            Highwinds  & 504  & 26   & 5\%               \\
            Edgecast   & 241  & 16   & 7\%               \\
            Incapsula  & 216  & 142  & \color{red}{66}\% \\
            Quantil    & 161  & 10   & 6\%               \\
            CDNetworks & 32   & 3    & 9\%               \\
            Limelight  & 30   & 5    & 17\%              \\
            \hline
        \end{tabular}
    }
    \hfill
    \subfloat[\label{tab:CategoryDist}\textbf{Website categories}]{
        \centering
        \begin{tabular}{>{\bfseries\scriptsize}l|c|c|c}
            % Retail & Internet & Business & Entertainment & News & Finance & Technology &
            % Education & Society & Travel & Science & Sports & Health & Reference & Government
            % & Recreation & Home
            \hline
            \scriptsize{\textbf{~~Category}} & \scriptsize{\textbf{\makecell{CDN-                           \\enabled}}} & \scriptsize{\textbf{\makecell{Password-\\exposed}}} & \scriptsize{\textbf{Percent}} \\
            \hline
            Retail                           & 304                                & 175 & \color{red}{58\%} \\
            Internet                         & 231                                & 69  & 30\%              \\
            Business                         & 225                                & 72  & 32\%              \\
            Entertain                        & 213                                & 76  & 36\%              \\
            News                             & 181                                & 62  & 34\%              \\
            Finance                          & 159                                & 60  & \color{red}{38}\% \\
            Technology                       & 155                                & 42  & 27\%              \\
            Education                        & 145                                & 14  & 10\%              \\
            Society                          & 99                                 & 31  & 31\%              \\
            Travel                           & 79                                 & 34  & \color{red}{43\%} \\
            Science                          & 50                                 & 18  & 36\%              \\
            Sports                           & 49                                 & 15  & 31\%              \\
            Health                           & 43                                 & 17  & \color{red}{40\%} \\
            Reference                        & 36                                 & 13  & 36\%              \\
            % Government                       & 16                                 & 3   & 19\% \\
            % Recreation                       & 16                                 & 7   & \color{red}{44\%} \\
            % Home                             & 9                                  & 6   & \color{red}{67\%} \\
            \hline
        \end{tabular}
    }
\end{table}

% \begin{figure*}[t]
% 	\centering
% 	\includegraphics[width=0.8\textwidth]{figures/CDNDist.eps}
% 	\caption{Distribution across CDN providers. The y-axis is in log scale. The
% 		number above the bar denotes the percentage of password-exposed websites in
% 		CDN-enabled websites.}\label{fig:CDNDist}
% \end{figure*}

% \begin{figure*}[t]
% 	\centering
% 	\includegraphics[width=0.8\textwidth]{figures/CategoryDist.eps}
% 	\caption{Distribution across website categories. The number above the bar
% 	denotes the percentage of password-exposed websites in CDN-enabled websites.}
% 	\label{fig:CategoryDist}
% \end{figure*}

\subsection{Distribution over CDN Providers}
We also consider how password-exposed websites are distributed among the CDN
providers. Table~\ref{tab:CDNDist} presents the number of password-exposed
websites in each CDN provider. As shown in the table, Cloudflare and Akamai are the
two most popular CDNs in the world, and they observe the most users' passwords
from their customers' requests. More than 40\% of Cloudflare's customer websites in our dataset share users' passwords to Cloudflare, and Akamai observes passwords from 25\% of its customers.
Besides, 66\% of websites that use Incapsula expose passwords to the CDN.
Some CDNs only observe a small fraction of sensitive traffic, such as Highwinds and Edgecast.

\mr{Compared to the other CDN providers, a much larger portion of Cloudflare and Incapsula customers are affected by password exposure. For Cloudflare, the reason may be the difference in request redirection methods. Cloudflare uses anycast for request redirection by default~\cite{AnycastCDN-IMC15}, while the other CDNs use DNS redirection~\cite{Akamai-SIGOPS10,Akamai-SGICOMMCCR15}. As discussed in ~\cite{InviCloak-CCS22}, to enable anycast redirection, a website needs to use Cloudflare as the DNS provider. Such a practice will transfer a website's all DNS records to Cloudflare DNS service, including the resolution to the domain of the login request (\eg DNS A record of \texttt{login.example.com}). Cloudflare will conduct anycast redirection for the transferred domains by default. Therefore, the login request is very likely to be terminated by Cloudflare. We verify this inference by checking the DNS provider of password-exposed websites using Cloudflare. We find that 63\% of websites that transferred their DNS providers to Cloudflare expose their passwords to Cloudflare, while 83\% of websites that use Cloudflare CDN service without transferring their DNS providers do not suffer the password exposure.}

\mr{As for Incapsula, such a high percentage (66\%) may originate from the dynamic content caching provided by Incapsula~\cite{DynamicContentCaching}. Such a service will cache the dynamic content for a short period to improve the performance of webpage loading, which is not enabled by the other CDN providers. Websites using Incapsula may employ this service to cache the dynamic content including the login responses, leading to password exposure.}

It is reasonable for websites to trust famous CDN providers and
employ their defense against attacks. However, it does not necessarily mean
users should also trust CDNs. From the users' perspective, they may be concerned about their
private data when it is shared with a third-party CDN. The
results also imply a risk of the single point failure of popular CDNs: a
malicious insider in a popular CDN may divulge the users' passwords of more than
40\% of its customer websites, leading to a large-scale user data leakage.

\mr{We reported our findings to three CDN providers, Cloudflare, Akamai, and Fastly. All of them replied to us. They acknowledged the implication of password exposure and claimed that they are trustworthy and will follow the privacy policy~\cite{CloudflareCompliance,FastlyPolicy} to secure customers' data.
Akamai also explained that they must terminate the TLS connections including those transmitting private data in order to provide protections such as WAF for customers.}

\subsection{Distribution over Website Categories}
We investigate the practice of exposing passwords among different website categories. We
collect the website category data from Alexa Top Sites by
Category~\cite{AlexaTopSites}. In 12,451 CDN-enabled websites, 2,010 of them can
be classified by the Alexa data. In our dataset, three categories (Government, Recreation, and Home) contain less than 20 CDN-enabled websites, so we consider the dataset is not representative enough for these three categories. Thus, we only use the rest of the 14 categories in our analysis in this section.
%We use these 2,010 sites for analysis in this section.

Table~\ref{tab:CategoryDist} presents the statistics of CDN usage and password
exposure across 14 website categories. As we can see, retail websites employ
most CDNs because they need to display many pictures of their
products, and CDNs notably accelerate the picture delivery . 
\mr{However, most retail websites (58\%) also expose passwords to CDNs. 
Besides retail websites, more than 40\% of
websites of travel and health expose users' passwords.} We note that a
large portion (38\%) of finance and health websites which are usually considered to
require sophisticated defense divulges users' passwords to CDNs. Moreover, education
websites have the least percentage of password exposure. \mr{Our results point out
that password exposure is prevalent within a wide range of categories, while retail, 
travel, and health are the most affected website categories.}

\begin{figure}[t]
	\centering
	\includegraphics[width=0.48\textwidth]{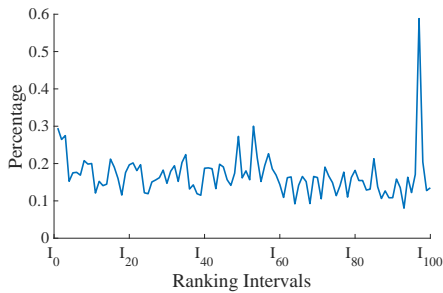}
	\caption{Percentages of password-encrypted websites among CDN-enabled websites across different ranking intervals. We divide 50K websites into 100 intervals. Each interval contains 500 websites. The x-axis ticks at every 20 intervals.}\label{fig:SharePercentEnc}
\end{figure}

% \begin{figure*}[tb]
% 	\centering
% 	\includegraphics[width=0.8\textwidth]{figures/CategoryDistEnc.eps}
% 	\caption{Distribution across website categories. The number above the bar
% 	denotes the percentage of password-encrypted websites in CDN-enabled
% 	websites.} \label{fig:CategoryDistEnc}
% \end{figure*}

\section{Countermeasures}~\label{Countermeasures} 
In this section, we first present the measurement of the countermeasures against password exposure used
by current websites. We also discuss possible countermeasures that websites and users can adopt.

\subsection{Client-side Encryption and CDN Bypassing}~\label{EncryptionAndBypassing}
In our measurement, we observe that some websites indeed adopt
client-side encryption discussed in \S~\ref{ExistingSolutions} to protect users'
passwords. For example, \texttt{baidu.com},
\texttt{dropbox.com}, and \texttt{chase.com} deliver public keys by their origin
servers. However, such a solution is rarely adopted by the websites. In our
measurement, if our framework submits the credentials but cannot find the
password in plain text or Base64 encoding in the login request, we consider that the website
encrypts the password. Since our framework may fail to login, we have an upper-bound estimation of the deployment of client-side
password encryption. Therefore, in our dataset, at most 2,057 (16.5\%) out of
12,451 CDN-enabled websites adopt such a solution. We call these websites
``password-encrypted websites''. This result demonstrates that password
encryption is a rare practice on the web.

We investigate the relationship between a website's ranking and password encryption
deployment. We used the same method and intervals in
Figure~\ref{fig:SharePercent}, and the results are shown in
Figure~\ref{fig:SharePercentEnc}. As we can see, even for the websites that rank top 1,500 ($I_0, I_1,$ and $I_2$), 
less than 30\% of them encrypt users' passwords. Nevertheless, when compared with
other websites with lower ranks, they have a relatively higher percentage of password encryption. 
However, an outstandingly high percentage exists around the intervals of quite
low rankings. We manually inspected websites located in that interval. We found 13 websites of all 20 password-encrypted websites 
are subdomains of \texttt{tmall.com} for different retailers, such as
\texttt{www.kfc.tmall.com} and \texttt{www.lenovo.tmall.com}. Once a user attempts to sign
into these subdomain sites, they all direct the user to \texttt{tmall.com}. This website is a top electronic shopping website, and it adopts password encryption.

% We also investigate the relationship between the website's category and password
% encryption. As shown in Figure~\ref{fig:CategoryDistEnc}, in most categories, less than 20\% of the websites encrypt passwords. Government websites have a relatively high percentage of password encryption. It may be so because government websites contain relatively more sensitive data, and thus they prefer password encryption. In contrast, a small percentage of websites in categories such as home, education, and reference adopts password encryption. Overall, this figure indicates that the password encryption practice is rare in most website categories.

\mr{We note that as a preliminary defense, client-side encryption can only defend against passive attackers as described in \S~\ref{ThreatModel}. However, our measurement shows that most websites including top ones cannot even prevent a passive attacker.}  If an active attacker exists, CDN bypassing can protect users' privacy, but it exposes origin servers' IP addresses and leave servers at the risk of DDoS. In our measurement,
we cannot verify whether the destination of a login request is the origin server through RDAP.
% because (1) Some websites or IP addresses do not contain an RDAP entry; (2) The RDAP result does not necessarily include the website's domain name, because some
% websites' organization names do not contain the domain names. For example,
% ``youtube.com'' uses ``Google LLC'' in its RDAP result; (3) A website may be
% hosted in a cloud provider, and the RDAP results only show the information of
% the cloud provider. For example, the RDAP result of ``quora.com'' is ``AWS''.
We leave the further measurement of CDN bypassing as future work.

\subsection{Possible Countermeasures}~\label{OtherCountermeasures}
Besides client-side encryption and CDN bypassing, Password Authenticated Key
Exchange (PAKE)~\cite{PAKE-EUROCRYPT00,PAKE} also prevents password exposure.
PAKE protocols, such as SRP~\cite{SRP-NDSS98} and
OPAQUE~\cite{OPAQUE-EUROCRYPT18}, authenticate users without the requirement of
revealing passwords in login requests. Moreover, it is proven to be secure during
login even when CDNs can launch active attacks. However, PAKE protocols
require trust on first use (TOFU), meaning that a secure channel is required
during account registration. Therefore, PAKE solves the password
exposure issue for web services that do not allow online registration. For
example, it can be used in banking industry, as users are required to open a bank account physically at branches.
Nevertheless, PAKE is almost never used by websites~\cite{PAKE}. The reason may
be the difficulty of understanding and implementing PAKE protocols for
developers. It may also be because developers usually trust third-party CDNs and are not aware of such a password exposure issue.

From the users' perspective, a user can use OAuth~\cite{RFC6749-OAuth} such as using
a Google account to sign in to other websites. Because leading tech companies such as
Google and Facebook have built their own CDNs, a user's password will not be
exposed to a third party during the login. However, more OAuth practices may lead
to a severe single-point failure if a user's password of the Google account is
leaked. Besides OAuth, users can also adopt two-factor authentication. Even
though two-factor authentication cannot prevent passwords from being exposed to
third-party CDNs, it prevents accounts from being compromised even when the
passwords are exposed to attackers.

These countermeasures can only protect users' passwords. However, users' private data stored on a website may also be divulged to a CDN during the transmission. As private data are much more
complicated and diverse than the passwords, developing countermeasures would
be harder. Thus, private data leakage may be much more prevalent than
password leakage. We leave the measurement of private data leakage as future
work.
\section{Discussion and Future Work}~\label{Discussion} 
Our measurement quantifies password exposure to CDNs and suggests potential security issues in current web ecosystem.
In this section, we provide suggestions to the security community, users, and 
the industry.

\emph{We need further research on the solutions.} As presented in
\S~\ref{ExistingSolutions}, the preliminary strategies of CDN bypassing and
client-side encryption can be easily deployed but contain vulnerabilities.
Proposed techniques such as Keyless
SSL~\cite{CloudflareKeylessSSL,AkamaiKeylessSSL,WASP}, certificate
delegation~\cite{HTTPSonCDN-SP14}, and mcTLS~\cite{mcTLS-SIGCOMM15} are
ineffective in preserving user privacy. 
The SGX-based solutions~\cite{mbTLS-CoNEXT17,Phoenix-Security20} can provide
comprehensive protection, but it is hard to be deployed on CDNs. 
InviCloak~\cite{InviCloak-CCS22} can achieve 
the goal of DDoS defense, privacy protection, and instant deployment simultaneously, but it disables the Web Application Firewall (WAF) of CDNs.
% and it is still unclear whether website developers is willing to adopt InviCloak.
Therefore, further research on this area is critical to a more secure Internet.

\emph{We recommend users adopt two-factor authentication.} 
% As discussed in \S~\ref{OtherCountermeasures}, signing into a website with existing
% accounts of leading tech companies such as Google and Facebook could be safer than creating a new account. 
Two-factor authentication provides additional protection for an account even when the password is stolen by a hacker. 
\mr{Adopting OAuth is debatable as it may lead to the single point of failure although it prevents password exposure as discussed in \S~\ref{OtherCountermeasures}.}
%We also encourage users to know about the security issue presented in this paper.

\emph{Websites should adopt preliminary defense.} The results shows that
many websites do not apply the minimal defense against password exposure. Despite the preliminary strategies
are vulnerable to some attacks, they provide basic protection for users' privacy.
Since it is acceptable to assume a passive CDNs in most cases, the
client-side encryption usually provides a sufficient protection. 

\emph{CDN providers should involve in developing and deploying advanced solutions.}
The widespread of Keyless SSL on Cloudflare demonstrates that a CDN provider
plays an important role in the security community~\cite{CloudflareKeylessSSL}. Cooperation from CDN providers can
validate researchers' ideas and advance further research. CDNs can also guide
their customers to deploy a defense mechanism.

This paper presents the preliminary results of password sharing to third-party CDNS. We propose the following directions as the future work.
\begin{enumerate}
    \item Augment the existing CDN discovery method to differentiate the hosting service and the CDN service of a cloud provider, as mentioned in \S~\ref{Method}.
    \item Quantify the adopted or available countermeasures besides the client-side encryption in websites, including CDN bypassing, OAuth, one-time password, two-factor authentication, etc, as mentioned in \S~\ref{Countermeasures}.
    \item Measure private data leakage in websites to understand the security impact of TLS private key sharing from users' perspectives, as mentioned in \S~\ref{Countermeasures}.
    %We can differentiate the private and public data by autenticated cookies in the requests.
    \item Survey the users and website developers to understand their awareness of private data leakage to thrid-party CDNs. Such a survey helps to figure out the reason why countermeasures are not widespread.
\end{enumerate}
\section{Related Work}~\label{RelatedWork}
\emph{Password security.} Password security has attracted attention from many
researchers. Lu~\et analyzed how websites deploy measures to prevent online
password cracking~\cite{Authentication-ACSAC18}. Wang~\et manually inspected 188
websites to characterize the login process and built an extension to inform
users of potential password leakage caused by the lack of
HTTPS~\cite{LoginSafetly-SIGCOMMWorshop11}. Acker~\et studied the security of
password input fields among the Alexa top 100K sites, and they found that
62.8\% of the websites with a login page are vulnerable to basic man-in-the-middle
attacks~\cite{LoginSecurity-SAC17}. Bonneau~\et surveyed the proposals for
replacing passwords and pointed out the difficulty of replacing
passwords~\cite{ReplacePassword-SP12}. Peng~\et explored how passwords are
spread after they are divulged by phishing sites~\cite{Phishing-AsiaCCS2019}. In
addition, many prior works investigated the prevalence of the password reuse
problem~\cite{PasswordHabit-CCS17,SecurityPractice-SOUPS15,PasswordReuse-SOUPS16,PasswordAttitudes-SOUPS10} and its countermeasures~\cite{PasswordReuse-NDSS19}.

\emph{CDN security.} Researchers have shown the existence of a wide range of
vulnerabilities in CDNs. Mirheidari~\et's measurement
shows that private data can be divulged by CDNs through web cache
deception~\cite{WCD,WCD-Security20,WCD-Security22}. Nguyen~\et presented an
attack of poisoning CDN cache with error pages, and five CDN services were
vulnerable to such an attack~\cite{CPDoS-CCS19}. Besides CDN cache, researchers also
presented approaches to disclosing the IP addresses of origin servers hidden
behind CDNs, demonstrating insufficient DDoS protection of
CDNs~\cite{BypassCBSP-CCS15,CDNDDoS-DSN18}. Moreover, attackers may utilize a
CDN to launch DoS to an origin server or to the CDN itself~\cite{CDNThreat-ESORICS09,ForwardingLoop-NDSS16,CDNJudo-NDSS20}.
% Triukose~\et presented an amplify method to launch DoS to an origin server through the CDN~\cite{CDNThreat-ESORICS09}. 
% The forwarding loop discovered by Chen~\et can
% lead to resource-consuming DoS to CDNs~\cite{ForwardingLoop-NDSS16}. 
% Guo introduced three attacks to break CDN DoS protection, including HTTP/2
% amplification, pre-post slow HTTP, and availability degradation~\cite{CDNJudo-NDSS20}. 
In addition, 
% Hao~\et's research demonstrated
% that attackers can hijack the DNS redirection used by a CDN to downgrade the
% content delivery performance~\cite{RedirectionHijacking-Security18}.
Durumeric~\et's measurement shows that the HTTPS interception on CDNs may
downgrade the TLS version or cipher suites and thus reduce connection
security~\cite{Interception-NDSS17}.

\emph{Solutions to TLS key sharing.} A line of research focuses on building
keyless CDNs. Cloudflare, Akamai, and Modadugu~\et proposed similar solutions
called ``Keyless SSL'',
respectively~\cite{CloudflareKeylessSSL,AkamaiKeylessSSL,WASP}. Certificate delegation~\cite{HTTPSonCDN-SP14} and
mcTLS~\cite{mcTLS-SIGCOMM15} enable a client to recognize the CDN as a
delegation of the website. Wei~\et~\cite{STYX-SoCC17} and
Ahmed~\et~\cite{Harpocrates-SEC18} adopted Trust Executive Environment (TEE) on
CDNs for private key management. However, these strategies only prevent the TLS
private key sharing, while users' private data are still visible to CDNs.
Phoenix~\cite{Phoenix-Security20} and mbTLS~\cite{mbTLS-CoNEXT17} extend TEE
solutions to fully protect users' private data. However, deploying TEE-based
solutions on CDNs may take a long time as it requires upgrades of hardware and
operating systems. InviCloak~\cite{InviCloak-CCS22} protects users' private data with
an additional encryption channel and low overhead, but its adoption by websites
in the future remains unclear.
\section{Conclusion}~\label{Conclusion}
In this paper, we conduct a large-scale measurement to quantify user password exposure to third-party CDNs in the web ecosystem. Our results show that 33.0\% of CDN-enabled websites expose users' passwords to the CDNs during the login procedures. 
\mr{Retail websites substantially benefit from CDNs but also tend to expose passwords to CDNs.} 
\mr{Besides, client-side password encryption is adopted by less than 17\% of websites, even though it is simple and effective to a certain extent.}
Overall, our results suggest that current websites excessively trust CDNs, leading to potential security issues when attackers exploit CDNs' vulnerabilities.
% As HTTPS and CDNs becomes more popular, we encourage further research on the privacy issues caused by HTTPS termination on CDNs. 
We publicly released the code to facilitate future research~\cite{GitHubCode-PAM23}. 

\subsubsection*{Acknowledgements.} We sincerely thank our shepherd Georgios Smaragdakis and anonymous reviewers for their helpful comments. This work is supported in part by the Duke CS+ summer research program and NSF award CNS-1901047.

\section*{Appendix}\label{Appendix}

\begin{mrlong}
    We present the detail of our auto-login framework in this section.
    For each web page, the framework applies four steps to the HTML elements:
    filtering, classifying, scoring, and submitting credentials. The framework first
    filters the elements based on tag names and locations. Then it uses keyword
    frequency as the features to classify filtered elements into three classes:
    login entrances, account inputs, and password inputs. In each class, it assigns a
    score to each element according to features extracted from the HTML code.
    Finally it fills and submits credentials if the login form is found, or it
    clicks on the login entrance to visit the login page. The elements to interact with are
    chosen by their scores in each class. The followings paragraphs introduce each step
    in detail.

    \begin{enumerate}
        \item \textbf{Filtering}: When the framework arrives at a page, it starts
              with filtering out elements that are considered irrelevant to login.
              Specifically, it selects elements containing one of the following tag
              names: ``input'', ``button'', ``label'', ``a'' and ``iframe''.
              %   As some websites may use other tag names for their login entrances, the
              %   framework also uses the CSS ``cursor'' property to select elements
              %   that change the cursor into a pointing hand when hovering over, which
              %   is a common feature for links. 
              To reduce element candidates, we assume
              that a login entrance or a login form should be shown within the area of
              one and a half of the viewport height from the top of a web page. The
              rationale of this assumption is that a website should place login
              elements at positions that are easily accessible to users.
        \item \textbf{Classifying}: To classify an element into the classes mentioned above,
              the framework extracts strings from HTML properties and the inner text of the
              element. It then splits strings into words by camel case and non-word
              characters.  It computes the frequencies of some keywords in the
              string. The keyword frequencies are regarded as a feature of the
              element. The framework classifies the element based on these features
              and heuristic rules. We manually select eleven keywords and construct
              rules for classification after examining Alexa top 100 sites. One example of the rules is that a login entrance should contain at least
              one of the keywords related to ``login'', ``account'', or ``email''.
              To improve the detection accuracy, we also apply some deprecation keywords   such as ``user guide'' and ``policy''.  An element is discarded if it contains any of the deprecation keywords.
        \item \textbf{Scoring}: While a website usually contains only one login
              entrance, the framework may classify multiple elements into the login
              class. Thus, our framework assigns scores to elements. For each
              element, the framework extracts other features besides keywords, such
              as the length of inner text and the visibility of element. The
              framework uses the features to assign a score to each element
              according to the rules we construct manually. For example, in the
              class of login entrance, a visible and interactive element receives a
              higher score than ones that are not. The frequency of a keyword in an
              element is also factored in the scores. Finally, the framework sorts
              elements in each class according to their scores.
        \item \textbf{Submitting Credentials}: If the framework obtains any input
              element in the account class or the password class, it fills each
              input element with credentials. Then it uses the keyboard signal, ENTER, to
              submit fake credentials. If no input field is detected, the framework
              clicks on the login element with the highest score and repeats the
              presented steps on the new web page to detect input fields. The framework collects the login request once it considers a credential submission happens.
    \end{enumerate}

    Overall, our framework uses heuristic rules to detect login entrances and input
    fields of credentials. We implement the framework by using Selenium
    WebDriver~\cite{Selenium} to control Chrome. We test our framework on 100
    random-selected websites of which 52  enable the login. The results show that our
    framework successfully submits credentials to 45 of 53 websites, meaning a recall
    of 84.9\%. The framework ignores all 47 websites without a login entrance,
    meaning a false positive of 0\%. The overall detection accuracy is
    (45+47)/100=92.0\%.

    \emph{Existing automatic login frameworks:} Browsers such as Chrome and Firefox can help users automatically fill in the credentials on some web pages.
    We do not use this function because it relies on the existence of the ``autocomplete''
    attribute in HTML elements, and thus it cannot handle the websites that do not
    enable this attribute in HTML. Besides the automation of browsers, Peng~\et
    implemented a framework to log into phishing websites
    automatically~\cite{Phishing-AsiaCCS2019}. Our framework can handle issues
    that are common in legitimate sites but rare in phishing sites, such as
    confusion caused by sign-up forms and pop-ups.
    Jonker~\et. also proposed a framework for post-login security
    analysis~\cite{AutoLogin-NDSSWorkshop20}. Our framework shares many similarities
    with theirs but adds the capability to operate in the presence of HTTP
    Authentication and reCAPTCHA.
\end{mrlong}


\bibliographystyle{splncs04}
\bibliography{main}

\begin{thebibliography}{10}
\providecommand{\url}[1]{\texttt{#1}}
\providecommand{\urlprefix}{URL }
\providecommand{\doi}[1]{https://doi.org/#1}

\bibitem{Akamai}
{Akamai} (2020), \url{https://www.akamai.com/}

\bibitem{AlexaTopSites}
{Alexa Top Sites} (2020), \url{https://www.alexa.com/topsites}

\bibitem{Cloudflare}
{Cloudflare} (2020), \url{https://www.cloudflare.com/}

\bibitem{Fastly}
{Fastly} (2020), \url{https://www.fastly.com/}

\bibitem{Selenium}
{SeleniumHQ Browser Automation} (2020), \url{https://www.selenium.dev/}

\bibitem{AJAX}
{AJAX} (2022), \url{https://developer.mozilla.org/en-US/docs/Web/Guide/AJAX}

\bibitem{CloudflareCompliance}
{Certifications and Compliance Resources} (2022),
  \url{https://www.cloudflare.com/trust-hub/compliance-resources/}

\bibitem{DynamicContentCaching}
{Global CDN and Optimizer - Introduction} (2022),
  \url{https://docs.imperva.com/bundle/cloud-application-security/page/introducing/global-cdn-optimizer.htm}

\bibitem{GitHubCode-PAM23}
{PAM2023-CDNPassword} (2022),
  \url{https://github.com/SHiftLin/PAM2023-CDNPassword}

\bibitem{FastlyPolicy}
{Security Measures} (2022),
  \url{https://docs.fastly.com/en/guides/security-measures}

\bibitem{Harpocrates-SEC18}
Ahmed, R., Zaheer, Z., Li, R., Ricci, R.: {Harpocrates: Giving Out Your Secrets
  and Keeping Them Too}. In: Proc. of IEEE/ACM Symposium on Edge Computing
  (SEC). pp. 103--114. IEEE (2018)

\bibitem{ReplacePassword-SP12}
Bonneau, J., Herley, C., Van~Oorschot, P.C., Stajano, F.: The quest to replace
  passwords: A framework for comparative evaluation of web authentication
  schemes. In: Proc. of S\&P. pp. 553--567. IEEE (2012)

\bibitem{PAKE-EUROCRYPT00}
Boyko, V., MacKenzie, P., Patel, S.: Provably secure password-authenticated key
  exchange using diffie-hellman. In: Proc. of EUROCRYPT. pp. 156--171. Springer
  (2000)

\bibitem{AnycastCDN-IMC15}
Calder, M., Flavel, A., Katz-Bassett, E., Mahajan, R., Padhye, J.: Analyzing
  the performance of an anycast cdn. In: Proc. of IMC. pp. 531--537 (2015)

\bibitem{KeySharing-CCS16}
Cangialosi, F., Chung, T., Choffnes, D., Levin, D., Maggs, B.M., Mislove, A.,
  Wilson, C.: {Measurement and Analysis of Private Key Sharing in the HTTPS
  Ecosystem}. In: Proc. of CCS. pp. 628--640. ACM (2016)

\bibitem{ForwardingLoop-NDSS16}
Chen, J., Zheng, X., Duan, H.X., Liang, J., Jiang, J., Li, K., Wan, T., Paxson,
  V.: Forwarding-loop attacks in content delivery networks. In: Proc. of NDSS.
  ISOC (2016)

\bibitem{Interception-NDSS17}
Durumeric, Z., Ma, Z., Springall, D., Barnes, R., Sullivan, N., Bursztein, E.,
  Bailey, M., Halderman, J.A., Paxson, V.: The security impact of https
  interception. In: Proc. of NDSS. ISOC (2017)

\bibitem{AkamaiKeylessSSL}
Gero, C.E., Shapiro, J.N., Burd, D.J.: {Terminating SSL Connections without
  Locally-Accessible Private Keys} (2013), u.S. Patents, No. 9,647,835

\bibitem{WCD}
Gil, O.: {Web Cache Deception Attack} (2017),
  \url{https://omergil.blogspot.com/2017/02/web-cache-deception-attack.html}

\bibitem{Protecting-Computer15}
Gillman, D., Lin, Y., Maggs, B., Sitaraman, R.K.: {Protecting Websites from
  Attack with Secure Delivery Networks}. Computer  \textbf{48}(4),  26--34
  (2015)

\bibitem{PAKE}
Green, M.: {Let's Talk About PAKE} (2018),
  \url{https://blog.cryptographyengineering.com/2018/10/19/lets-talk-about-pake/}

\bibitem{CDNAbuse-SRDS18}
Guo, R., Chen, J., Liu, B., Zhang, J., Zhang, C., Duan, H., Wan, T., Jiang, J.,
  Hao, S., Jia, Y.: Abusing cdns for fun and profit: Security issues in cdns'
  origin validation. In: Proc. of SRDS. pp. 1--10. IEEE (2018)

\bibitem{CDNJudo-NDSS20}
Guo, R., Li, W., Liu, B., Hao, S., Zhang, J., Duan, H., Shen, K., Chen, J.,
  Liu, Y.: Cdn judo: Breaking the cdn dos protection with itself. In: Proc. of
  NDSS. ISOC (2020)

\bibitem{SQLInjection-ISSSE06}
Halfond, W.G., Viegas, J., Orso, A., et~al.: {A Classification of SQL-Injection
  Attacks and Countermeasures}. In: Proc. of International Symposium on Secure
  Software Engineering. vol.~1, pp. 13--15. IEEE (2006)

\bibitem{RFC6749-OAuth}
Hardt, D.: The oauth 2.0 authorization framework. Internet Engineering Task
  Force (IETF)  (2012)

\bibitem{Phoenix-Security20}
Herwig, S., Garman, C., Levin, D.: {Achieving Keyless CDNs with Conclaves}. In:
  Proc. of Security Symposium. pp. 735--751. USENIX (2020)

\bibitem{CDN-IMC08}
Huang, C., Wang, A., Li, J., Ross, K.W.: Measuring and evaluating large-scale
  cdns. In: Prof. of IMC. pp. 15--29. ACM (2008)

\bibitem{SecurityPractice-SOUPS15}
Ion, I., Reeder, R., Consolvo, S.: ``... no one can hack my mind'': Comparing
  expert and non-expert security practices. In: Proc. of SOUPS. pp. 327--346.
  USENIX (2015)

\bibitem{OPAQUE-EUROCRYPT18}
Jarecki, S., Krawczyk, H., Xu, J.: Opaque: an asymmetric pake protocol secure
  against pre-computation attacks. In: Proc. of EUROCRYPT. pp. 456--486.
  Springer (2018)

\bibitem{CDNDDoS-DSN18}
Jin, L., Hao, S., Wang, H., Cotton, C.: Your remnant tells secret: Residual
  resolution in ddos protection services. In: Proc. of DSN. pp. 362--373. IEEE
  (2018)

\bibitem{AutoLogin-NDSSWorkshop20}
Jonker, H., Karsch, S., Krumnow, B., Sleegers, M.: Shepherd: A generic approach
  to automating website login. In: Proc. of NDSS Workshop on Measurements,
  Attacks, and Defenses for the Web. ISOC (2021)

\bibitem{CDN-IMC01}
Krishnamurthy, B., Wills, C., Zhang, Y.: On the use and performance of content
  distribution networks. In: Proc. of IMC. pp. 169--182. ACM (2001)

\bibitem{CDNMeasure-BSThesis2017}
Levy, A.: CDNs and Privacy Threats: A Measurement Study. Ph.D. thesis,
  Princeton University (2017)

\bibitem{HTTPSonCDN-SP14}
Liang, J., Jiang, J., Duan, H., Li, K., Wan, T., Wu, J.: {When HTTPS meets CDN:
  A Case of Authentication in Delegated Service}. In: Proc. of S\&P. pp.
  67--82. IEEE (2014)

\bibitem{InviCloak-CCS22}
Lin, S., Xin, R., Goel, A., Yang, X.: {InviCloak: An End-to-End Approach to
  Privacy and Performance in Web Content Distribution}. In: Proc. of CCS. ACM
  (2022)

\bibitem{Authentication-ACSAC18}
Lu, B., Zhang, X., Ling, Z., Zhang, Y., Lin, Z.: A measurement study of
  authentication rate-limiting mechanisms of modern websites. In: Proc. of
  ACSAC. pp. 89--100 (2018)

\bibitem{Akamai-SGICOMMCCR15}
Maggs, B.M., Sitaraman, R.K.: {Algorithmic Nuggets in Content Delivery}. ACM
  SIGCOMM CCR  \textbf{45}(3),  52--66 (2015)

\bibitem{WCD-Security20}
Mirheidari, S.A., Arshad, S., Onarlioglu, K., Crispo, B., Kirda, E., Robertson,
  W.: Cached and confused: Web cache deception in the wild. In: Proc. of
  Security Symposium. pp. 665--682. USENIX (2020)

\bibitem{WCD-Security22}
Mirheidari, S.A., Golinelli, M., Onarlioglu, K., Kirda, E., Crispo, B.: Web
  cache deception escalates! In: Proc. of Security Symposium. pp. 179--196.
  USENIX, Boston, MA (2022)

\bibitem{WASP}
Modadugu, N., Goh, E.J.: {The Design and Implementation of WASP: A Wide-Area
  Secure Proxy}. Tech. rep., Stanford University (2002)

\bibitem{mbTLS-CoNEXT17}
Naylor, D., Li, R., Gkantsidis, C., Karagiannis, T., Steenkiste, P.: And then
  there were more: Secure communication for more than two parties. In: Proc. of
  CoNEXT. pp. 88--100 (2017)

\bibitem{mcTLS-SIGCOMM15}
Naylor, D., Schomp, K., Varvello, M., Leontiadis, I., Blackburn, J., Lopez,
  D.R., Papagiannaki, K., Rodriguez~Rodriguez, P., Steenkiste, P.:
  Multi-context tls (mctls): Enabling secure in-network functionality in tls.
  In: Proc. of SIGCOMM. pp. 199--212. ACM (2015)

\bibitem{RFC7482-RDAP}
Newton, A., Hollenbeck, S.: Rfc7482: Registration data access protocol (rdap)
  query format. Internet Engineering Task Force (IETF)  (2015)

\bibitem{CPDoS-CCS19}
Nguyen, H.V., Iacono, L.L., Federrath, H.: Your cache has fallen:
  Cache-poisoned denial-of-service attack. In: Proc. of CCS. pp. 1915--1936.
  ACM (2019)

\bibitem{Akamai-SIGOPS10}
Nygren, E., Sitaraman, R.K., Sun, J.: {The Akamai Network: a Platform for
  High-Performance Internet Applications}. SIGOPS OSR  \textbf{44}(3),  2--19
  (2010)

\bibitem{PasswordHabit-CCS17}
Pearman, S., Thomas, J., Naeini, P.E., Habib, H., Bauer, L., Christin, N.,
  Cranor, L.F., Egelman, S., Forget, A.: Let's go in for a closer look:
  Observing passwords in their natural habitat. In: Proc. of CCS. pp. 295--310.
  ACM (2017)

\bibitem{Phishing-AsiaCCS2019}
Peng, P., Xu, C., Quinn, L., Hu, H., Viswanath, B., Wang, G.: What happens
  after you leak your password: Understanding credential sharing on phishing
  sites. In: Proc. of AsiaCCS. pp. 181--192 (2019)

\bibitem{LeakyForm-Security22}
Senol, A., Acar, G., Humbert, M., Borgesius, F.Z.: {Leaky Forms: A Study of
  Email and Password Exfiltration Before Form Submission}. In: Proc. of
  Security Symposium. pp. 1813--1830. USENIX, Boston, MA (2022)

\bibitem{PasswordAttitudes-SOUPS10}
Shay, R., Komanduri, S., Kelley, P.G., Leon, P.G., Mazurek, M.L., Bauer, L.,
  Christin, N., Cranor, L.F.: Encountering stronger password requirements: User
  attitudes and behaviors. In: Proc. of SOUPS. pp. 1--20. USENIX (2010)

\bibitem{Akamai5YearsDDoS}
Sparling, C.: {5 Years of Fighting DDoS with the Power of Akamai} (2019),
  \url{https://blogs.akamai.com/2019/07/5-years-of-fighting-ddos-with-the-power-of-akamai.html}

\bibitem{CloudflareKeylessSSL}
Sullivan, N.: {Keyless SSL: The Nitty Gritty Technical Details} (2014),
  \url{https://blog.cloudflare.com/keyless-ssl-the-nitty-gritty-technical-details/}

\bibitem{CDNThreat-ESORICS09}
Triukose, S., Al-Qudah, Z., Rabinovich, M.: Content delivery networks:
  Protection or threat? In: Proc. of ESORICS. pp. 371--389. Springer (2009)

\bibitem{LoginSecurity-SAC17}
Van~Acker, S., Hausknecht, D., Sabelfeld, A.: Measuring login webpage security.
  In: Proc. of the SAC. pp. 1753--1760. ACM (2017)

\bibitem{BypassCBSP-CCS15}
Vissers, T., Van~Goethem, T., Joosen, W., Nikiforakis, N.: Maneuvering around
  clouds: Bypassing cloud-based security providers. In: Proc. of CCS. pp.
  1530--1541. ACM (2015)

\bibitem{PasswordReuse-NDSS19}
Wang, K.C., Reiter, M.K.: How to end password reuse on the web. In: Proc. of
  NDSS. ISOC (2019)

\bibitem{LoginSafetly-SIGCOMMWorshop11}
Wang, X.S., Choffnes, D., Gage~Kelley, P., Greenstein, B., Wetherall, D.:
  Measuring and predicting web login safety. In: Proc. of SIGCOMM Workshop on
  Measurements up the Stack. pp. 55--60 (2011)

\bibitem{PasswordReuse-SOUPS16}
Wash, R., Rader, E., Berman, R., Wellmer, Z.: Understanding password choices:
  How frequently entered passwords are re-used across websites. In: Proc. of
  SOUPS. pp. 175--188. USENIX (2016)

\bibitem{STYX-SoCC17}
Wei, C., Li, J., Li, W., Yu, P., Guan, H.: {STYX: A Trusted and Accelerated
  Hierarchical SSL Key Management and Distribution System for Cloud Based CDN
  Application}. In: Proc. of SoCC. pp. 201--213. ACM (2017)

\bibitem{XSS-ESORICS11}
Weinberger, J., Saxena, P., Akhawe, D., Finifter, M., Shin, R., Song, D.: {A
  Systematic Analysis of XSS Sanitization in Web Application Frameworks}. In:
  ESORICS. pp. 150--171. Springer (2011)

\bibitem{SRP-NDSS98}
Wu, T.D., et~al.: The secure remote password protocol. In: NDSS. vol.~98, pp.
  97--111. Citeseer (1998)

\end{thebibliography}
\end{document}